\documentclass[aps,prd,preprint]{revtex4-1}
\usepackage[utf8]{inputenc}
\usepackage[pdftex]{graphicx}
\usepackage{amsmath,amssymb}
\usepackage{mathtools}
\usepackage{physics}
\usepackage{hyperref}
\hypersetup{
    colorlinks,
    citecolor=blue,
    filecolor=blue,
    linkcolor=blue,
    urlcolor=blue
}
\def\beq{\begin{equation}}
\def\eeq{\end{equation}}
\def\bea{\begin{eqnarray}}
\def\eea{\end{eqnarray}}

\renewcommand{\v}[1]{ \ensuremath{ {\bm{#1}} }}                  
\parskip=\itemsep               

\begin{document}

\title{Eikonal Expansion for Classical Gluon Fields: from Weak to Strong.}
\date{\today}
\author{Shane Brown, Alex DeBrizzi and Alex Kovner}

\affiliation{ Physics Department, University of Connecticut, Storrs, CT 06269, USA}
\begin{abstract}
{We develop a formal procedure of eikonal expansion for classical fields in gluodynamics. We show that in the weak field limit the expansion is straightforward and determines the eikonal components of vector potentials in terms of eikonal components of color currents via classical equations motion. We demonstrate that the recently proposed Born-Oppenheimer high energy evolution yields the ground state wave function interpretable in terms of the classical fields. The relevant classical field is indeed consistent with the  eikonal expansion, and contains the leading eikonal and subeikonal components. For strong fields we find that the naive eikonal expansion is inconsistent, as nonlilearities in equations of motion lead to mixing of eikonal orders. Thus the effect of subeikonal components on the leading eikonal component is not suppressed by the eikonality parameter. We show that this problem can be rectified by classically "integrating out" high longitudional momentum components of gluon fields. This leads to a classical action for the low momentum components with small self interaction, but strongly renorlamized currents, which contain contribution due to high momentum modes.}
\end{abstract}

\maketitle

\section{Introduction}

The approach involving quasi classical gluon fields has been very fruitful in formulating the high energy evolution in QCD. The JIMWLK evolution equation \cite{jimwlk,jimwlk1, jimwlk2, jimwlk3, jimwlk4, jimwlk5, jimwlk6, jimwlk7} has been derived starting with the classical solution for soft gluon fields in the background of a large source due to the valence degrees of freedom \cite{mv, mv1, mv2}. The BFKL equation \cite{bfkl, bfkl1, bfkl2}, albeit not originally derived in this way, can be formulated as  a limit of JIMWLK evolution where the color sources, and therefore the color fields are assumed to be weak. The form of the classical gluon fields expressed in terms of the color charge density  is preserved at any step of the evolution. This feature allows one to understand the JIMWLK equation  as evolving the distribution of color charge sources with energy due to "integrating out" softer and softer gluons.

In recent years two distinct generalizations of high energy evolution have been put forward. On one hand, the calculation of subeikonal corrections has been vigorously pursued by several groups in the context of calculating a variety of transverse dependent distributions (TMD's).  These are necessary since most of TMD's, including those determining components of a hadron spin evolve nontrivially only at next-to-leading eikonal order\cite{yuri1, yuri2, yuri}, \cite{tolga1, tolga2, tolga}.  On the other hand, the Born - Oppenheimer (BO) approach to the evolution was proposed in \cite{BO1, BO2} with the aim of unifying the low x (BFKL/JIMWLK) and high $Q^2$ (DGLAP) evolutions. As opposed to the original JIMWLK derivation, none of these generalizations directly use solutions of classical QCD equations of motion, albeit some elements of such a setup were examined in \cite{florian, florian1}. 

It would be very interesting to draw a more direct conceptual analogy between these  and JIMWLK, and see if these calculations can be formulated in a way that utilizes classical solutions to the QCD equations of motion. If this were the case, and if one could identify a finite number of sources (possibly beyond the color charge density which drives JIMWLK) which remain the only relevant ones at arbitrary energy, one can hope to formulate these calculations in terms of the flow of the joint distribution of these sources with energy, similarly to the JIMWLK evolution.
Such an approach would go along the  lines of \cite{florian}, where a weight functional for the leading eikonal term and the first subeikonal correction to the color current was constructed in analogy to the McLerran-Venugopalan model \cite{mv}, which is used heavily as an initial condition for CGC calculations \cite{mvincgc}.

Recently, a similar program based on quasi classical gluon fields as a fundamental starting point has been initiated in \cite{ming, ming1, ming2} in the context of subeikonal evolution. Ref. \cite{ming}  suggested an expansion in eikonality as the proper approach to finding subeikonal corrections to solutions of the classical Yang-Mills equations of motion (EOM). One purpose of the present paper is to sharpen the considerations of \cite{ming} by deriving a controlled and improvable expansion in eikonality. The parameter of this expansion, $\xi$ is the inverse value of the Lorentz $\gamma$-factor which defines the boost of the hadron from its rest frame to the frame in which it has large energy. In Section II we  derive this expansion and solve the classical equation of motion for the gluon fields to first subleading order in $1/\gamma$. Our solutions are slightly different from the ones given in \cite{ming}, and we discuss these differences. The bulk of the discussion in Section II is devoted to the weak field limit, i.e. we assume that the color sources/currents are parametrically small. However we also sharpen the interesting observation made in \cite{ming} that when the color sources are parametrically large, starting from the first subeikonal order, a consistent eikonal expansion seems to require  that the first order correction in eikonality also carry an additional power of $\alpha_s$ relative to the leading order field.

In Section III we turn to the Born-Oppenheimer approach. In \cite{BO1} the authors derived the wave function of fast gluon modes in the background of slow modes in the weak field limit. This wave function has the form of coherent state which hints at its possible classical origin. We show here that this is indeed the case. The calculation in \cite{BO1} can be understood in terms of. the solution of the classical equation of motion for fast gluon modes in the background of currents composed of the slow modes. However, as opposed to the leading eikonal approximation, where the only nonvanishing component of the color current is the one along the light cone direction, the BO solution is sourced also by the spatial components of the current. In addition the classical solution that corresponds to the BO wave function found in \cite{BO1} is not completely static, but rather "quasi static" - i.e. it takes into account the leading time dependence of the classical currents. This time dependence is responsible for nonvanishing of some of the subeikonal components in the solution.

In Sections IV and V we return to the question of the eikonal expansion for strong color fields. More carefully examining these equations we find that the naive eikonal expansion for strong sources is inconsistent in the following sense. Starting from the first subeikonal order, the EOM contain UV divergencies in the coefficients of terms that couple field components of different eikonality. This includes coupling of the leading eikonal component with all the subleading ones. These UV divergent terms are formally of higher order in the eikonality parameter $\xi$ and are always neglected in leading order calculations. However since the UV divergence arises from integration over longitudinal momentum, the only natural way to regulate this divergence is to introduce a UV momentum cutoff of the order $\Lambda\sim \mu/\xi$, with $\mu\sim \Lambda_{QCD}$ being a typical hadronic scale. Physically this corresponds to taking the longitudinal width of the shock wave to scale as $1/\gamma$. With such a regulator, the bare UV divergent coefficients become finite, but acquire extra factors of $1/\xi$ that change their eikonal counting. As a result the regulated EOM couple the leading eikonal mode with subleading modes already at {\bf leading} order. Thus one cannot solve the equations using the standard perturbative approach in which one first neglects subleading eikonal components, solves for the leading one and then includes the effects of the others perturbatively in $\xi$.

We show that the reason this problem arises is that in general the eikonal expansion is only valid at low longitudinal momentum $p^+$, while nonlinear terms which are local in coordinate space, require integration over all values of $p^+$.
Thus the way to resolve this problem is to expand eikonally only the low longitudinal momentum part of the vector potential. The high longitudinal momentum modes have to be treated as a background that couples to the low modes. One must therefore introduce a separation (factorization) scale $\Lambda=x\mu/\xi$ to separate the modes, with $x$ being an analog of Bjorken $x$ in the classical calculation. We show that for $x\ll 1$ the dynamics of low momentum modes is free of the problem mentioned above, in the sense that although the coupling between lower and higher eikonal components of the low momentum field exists, it is proportional to positive powers of $x$ and thus can be treated perturbatively. The price to pay is that the currents that source the low momentum fields undergo a "renormalization", i.e. must include the contribution of the high momentum modes in addition to the bare ("valence") currents. We demonstrate this for calculations in Lorenz gauge in Section IV and in Light cone gauge in Section V. This entirely classical procedure is analogous to the renormalization group approach to quantum JIMWLK and BFKL evolution\cite{jimwlk}, where the integration of high momentum modes leads to the evolution of color charge densities. However, as opposed to the quantum flow of currents, which is logarithmic, the classical "renormalization" is powerlike.

Finally Section VI is devoted to discussion of our results.

\section{Eikonal expansion in the weak field limit}
\subsection{Expansion in powers of $1/\gamma$.}
We start with the formal derivation of the expansion in the eikonality parameter. First consider a hadron at rest. We think of it as an (ensemble of) gluon field configurations, that are solutions of the classical EOM in the presence of color currents $j^\mu(x^+,x^-,x_\perp)$. The color currents are due to the "valence" particles. Naturally these valence degrees of freedom may be identified with quarks; however, their exact nature is of no importance except for the fact that they are not resolved in our approximation and that only the currents produced by them matter since they source the gluon fields. The color currents are regular in the rest frame. There they have finite support in a region of space on the order of hadronic size $\sim \Lambda^{-3}_{QCD}$, and fluctuate on the hadronic time scale, again $\sim \Lambda_{QCD}^{-1}$. We therefore assume that the spatial derivatives as well as moments of the currents are finite so that all quantities defined below are finite numbers with dimensions given by appropriate powers of $\Lambda_{QCD}$.

For the setup of the eikonal expansion that we have in mind, it is useful to define the following quantities, which (with a slight abuse of language) we refer to as "eikonal moments":
\begin{eqnarray}\label{eq:jmn}
&&j_{00}^\mu(x_\perp)\equiv \int dx^-j^\mu(x^+=0, x^-,x_\perp)\nonumber\\
&&j_{10}^\mu(x_\perp)\equiv \int dx^-\partial_+j^\mu(x^+, x^-,x_\perp)|_{x^+=0}\nonumber\\
&&j_{01}^\mu(x_\perp)\equiv -\int dx^-x^-j^\mu(x^+=0, x^-,x_\perp)\nonumber\\
&& j_{11}^\mu(x_\perp)\equiv -\int dx^-x^-\partial_+j^\mu(x^+, x^-,x_\perp)|_{x^+=0}\nonumber\\
&&...
\end{eqnarray}
For simplicity of notation we omit here and below the color index on currents and fields as long as we consider linear relations which hold for each color component separately. 

The gauge potentials  $a^\mu$ sourced by these currents also have finite derivatives and moments $a^\mu_{mn}(x_\perp)$, which we define similarly to \eqref{eq:jmn}.

Consider now boosting the hadron with the (large) boost parameter $\omega\gg 1$ to a frame in which it has high energy. Under this boost the currents and gauge potential transform as
    \begin{equation}\label{boost}
        \begin{aligned}
            A^+ (x^+,x^-,x_\perp) & = \xi^{-1}a^+(\xi x^+,\xi^{-1}x^-,x_\perp) \\
            A^i (x^+,x^-,x_\perp) & = \xi^0 a^i(\xi x^+,\xi^{-1}x^-,x_\perp) \\
            A^-(x^+,x^-,x_\perp) & = \xi a^-(\xi x^+,\xi^{-1}x^-,x_\perp)
        \end{aligned}
    \end{equation}
where $\xi\equiv e^{-\omega}=1/\gamma\ll 1$. 

Importantly, here we assume that the vector potentials are defined in a Lorentz covariant gauge, so that they transform as components of a proper Lorentz vector. We will consider the eikonal expansion in the ight cone gauge in the later sections.

Similarly the color current components in the boosted system are
 \begin{equation}\label{boost1}
        \begin{aligned}
            J^+ (x^+,x^-,x_\perp) & = \xi^{-1}j^+(\xi x^+,\xi^{-1}x^-,x_\perp) \\
            J^i (x^+,x^-,x_\perp) & = \xi^0 j^i(\xi x^+,\xi^{-1}x^-,x_\perp) \\
            J^-(x^+,x^-,x_\perp) & = \xi j^-(\xi x^+,\xi^{-1}x^-,x_\perp)
        \end{aligned}
    \end{equation}
    In the boosted system $A^+$ and $J^+$ are the largest components of the fields and currents respectively as $\xi\rightarrow 0$. To have access to corrections to this eikonal limit we expand the fields in powers of $\xi$. Due to \eqref{boost}, \eqref{boost1}, the expansion in powers of $\xi$ also prescribes the dependence on the coordinates $x^+$ and $x^-$. 
    
    In terms of $x^+$ it obviously corresponds to a Taylor expansion, as every power of $x^+$ comes with a power of $\xi$. In terms of $x^-$ on the other hand, the expansion in $\xi$ is equivalent to the expansion in derivatives $\partial^{(n)}\delta(x^-)$. To see this consider a function of a single variable
    \begin{equation}\label{Ff}
    F(x)=f(\xi^{-1}x)
    \end{equation}
where $f(y)$ is a regular function with finite support. An expansion in powers of derivatives of the $\delta$-function reads
\begin{equation}
F(x)=\delta(x)F_0+\delta'(x)F_1+\delta^{''}(x) F_2+...
\end{equation}
where
\begin{equation}
F_0=\int dx F(x); \ \ \ F_1=-\int dx xF(x); \ \ \ \ F_2=\frac{1}{2}\int dx x^2 F(x) \ \ \ \ ...
\end{equation}
Using \eqref{Ff} we have
\begin{equation}
F_0=\xi\int d yf(y)\equiv \xi f_0; \ \ \ F_1=-\xi^2\int dy y f(y)\equiv \xi^2 f_1 \ \ \ \ \ ...
\end{equation}
and thus
\begin{equation}
F(x)=\xi \delta(x)f_0+\xi^2\delta'(x)f_1+\xi^3\delta^{''}(x) f_2+...=\sum_{n=0}^\infty\frac{ (-1)^n}{n!}\xi^n\delta^{(n)}(x)[\int y^nf(y)dy]
\end{equation}
With this understanding we write the expansion of gluon fields and color currents (in the boosted frame)  in $\xi$ as

\begin{equation}
    \begin{aligned}
        &J^+(x^+,x^-,x_\perp)=\delta(x^-)j^+_{00}+\xi\left[x^+\delta(x^-)j^+_{10}+\delta'(x^-)j^+_{01}\right]+...\\
        &J^i(x^+,x^-,x_\perp)=\xi\delta(x^-)j^i_{00}+...\\
        &J^-(x^+,x^-,x_\perp)=\xi^2\delta(x^-)j^-_{00}+...
    \end{aligned}
\end{equation}
and
\begin{equation}
    \begin{aligned}
        &A^+(x^+,x^-,x_\perp)=\delta(x^-)a^+_{00}+\xi\left[x^+\delta(x^-)a^+_{10}+\delta'(x^-)a^+_{01}\right]+...\\
        &A^i(x^+,x^-,x_\perp)=\xi\delta(x^-)a^i_{00}+...\\
        &A^-(x^+,x^-,x_\perp)=\xi^2\delta(x^-)a^-_{00}+...
    \end{aligned}
\end{equation}
where all the coefficients in the expansion (the "eikonal moments") are finite and are determined by the currents and fields in the hadron rest frame.

The fields and currents are of course related through the solution of EOM. Given the eikonal expansion of currents and potentials one can attempt to formulate the equations of motion directly as equations for the eikonal components $a^\mu_{mn}(x_\perp)$. Such equations would be the extension of the standard eikonal approximation, which neglects all $a^\mu_{mn}(x_\perp)$ except $a^+_{00}(x_\perp)$, which in turn is expressed in terms of $j^+_{00}(x_\perp)$. 
If one is able to solve the equations for $a^\mu_{mn}(x_\perp)$ as a perturbative expansion in $\xi$, this procedure  would define a consistent eikonal expansion where higher subeikonal corrections to gluon field arise in higher orders in this perturbative expansion.  

The main goal of this paper is indeed to understand to what extent such a program is possible. We start our consideration with a weak field limit, assuming that all fields and currents are small.

\subsection{Solution to Linearized EOM in Lorenz gauge}
To determine the classical vector potentials in the boosted frame we must solve the QCD EOM:
    \begin{equation}\label{eom}
        D_\nu F^{\nu \mu}=J^\mu
    \end{equation}
     One can either solve the equations in the hadron rest frame, and then boost the solution, or equivalently solve directly in the boosted frame. The eikonal expansion is formulated directly in the boosted frame in the hope that finding the solution is simpler in such a formulation. We therefore work directly in the boosted frame.
    
 We start by assuming that all the currents and fields are weak, with the smallness determined by the QCD coupling constant, $J^\mu\sim g$.  In the following we keep only the first two orders in the eikonal expansion for simplicity. 
    
    Expanding \eqref{eom} to linear order in the fields, the first two terms in the eikonal expansion are given by
    \begin{equation}\label{sol1}
        \begin{aligned}
            \xi^0&: a^+_{00}  = -\frac{1}{\partial_\perp^2}j^+_{00}\\ 
            \xi^1&: a^+_{10}  = -\frac{1}{\partial_\perp^2}j^+_{10} \\
            \xi^1&: a^+_{01}  = -\frac{1}{\partial_\perp^2}\left( j^+_{01}-a^+_{10}-\partial^ja^j_{00}\right)\\
            \xi^1&: a^i_{00} -\frac{\partial^i\partial^j}{\partial_\perp^2}a^j_{00}+\frac{1}{\partial_\perp^2}\partial^ia^+_{10} = -\frac{1}{\partial_\perp^2}j^i_{00}        \end{aligned}
    \end{equation}

Note that the color current cannot be arbitrary, but has to satisfy the covariant conservation equation
\begin{equation}
D^{\mu\nu}J_\nu=0
\end{equation}
To order $\xi$ this reads
    \begin{equation}
        j^+_{10} = \partial^ij^i_{00}\ .
    \end{equation}
    Indeed, examining \eqref{sol1} we find that this is a necessary condition for the solution to  be consistent. 
    
    We also note that \eqref{sol1} does not determine the longitudinal component of $a^i_{00}$.
We can thus consistently impose the Lorenz gauge condition
\begin{equation}
\partial^\mu A_\mu=0
\end{equation}
which to order $\xi$ reads
 \begin{equation}
        a^+_{10} = \partial^ia^i_{00}\ .
    \end{equation}
 Using this condition explicitly, \eqref{sol1} is simplified
    \begin{equation}\label{sol2}
        \begin{aligned}
            \xi^0: a^+_{00} & = -\frac{1}{\partial_\perp^2}j^+_{00}\\ 
            \xi^1: a^+_{10} & = -\frac{1}{\partial_\perp^2}j^+_{10} \\
            \xi^1: a^+_{01} & = -\frac{1}{\partial_\perp^2}\left( j^+_{01}+2\frac{1}{\partial_\perp^2}j^+_{10}\right)\\
            \xi^1: a^i_{00} & = -\frac{1}{\partial_\perp^2}j^i_{00} 
        \end{aligned}
    \end{equation}
This determines the eikonal moments up to order $\xi$ in terms of the appropriate compoments of the color currents.
   
    Although here we have only kept the first two orders in the expansion, it is quite clear from this derivation that keeping more orders does not cause any difficulties. At each order the eikonal field is determined by the appropriate eikonal component of the color current. Thus we find that in the weak field limit the eikonal expansion can be consistently extended to arbitrary order.
    
    As we will see, the situation becomes considerably more complicated if the fields are not weak. We will postpone this discussion to Sections IV and V.

  \subsection{Comparison with solutions of \cite{ming}}
    
   Solutions of the classical equations in the eikonal expansion have been found in \cite{ming}. The procedure employed in  \cite{ming} is however somewhat different, since the fields are only expanded in powers of $x^+$, but are kept as nontrivial functions of $x^-$. 
    The solution  of \cite{ming} reads
    \begin{eqnarray}\label{solming}
   && \mathcal{A}_{-1}^+=-\frac{1}{\partial_\perp}J^+_{-1}\\
    &&\mathcal{A}^+_0=-\frac{1}{\partial^2_\perp}\left[2\partial_iD^+A_0^i+J^+_0\right]\nonumber\\
    &&\mathcal{A}^i_0=-\frac{1}{\partial^2_\perp}J_0^i\nonumber
    \end{eqnarray}
    where the subscripts correspond to the eikonal orders in the expansion of \cite{ming} and the covariant derivative $D^+$ is defined in the background of $\mathcal{A}_{-1}^+$. 
    
    The highest eikonal component directly corresponds to our definition $\mathcal A^+_{-1}=\delta(x^-)a^+_{00}$ and the solution \eqref{solming} is clearly equivalent to \eqref{sol2}. 
    
    The proper eikonal expansion at higher orders requires expanding \eqref{solming} in powers of $\xi$.  In addition, for the weak fields considered in this section the nonlinear term in \eqref{solming} is negligible. 
    
    To first nontrivial order in $\xi$, the fields in \eqref{solming} should be expanded as 
    \begin{equation}
    \mathcal{A}_0^+=x^+\delta(x^-)a^+_{10}+\delta'(x^-)a^+_{01}
    \end{equation}
    and so on.  Using the Lorentz gauge condition, we find that the linear terms in \eqref{solming} reduce exactly to the solutions \eqref{sol2}. Thus the solution of \cite{ming} is equivalent to the fully eikonally expanded solution we found above.
    
    \subsection{Eikonal expansion in the light cone (LC) gauge}
    In the context of JIMWLK evolution and in general in light cone perturbation theory it is convenient to use the light cone gauge $A^+=0$. 
In the light cone gauge the components of the vector potential do not transform under boost according to \eqref{boost}. The gauge condition $A^+=0$ is not invariant under Lorentz transformations, and thus a Lorentz transformation has to be accompanied by "regauging". The vector potentials in this gauge are therefore  not proper vectors.  One therefore cannot use the eikonal expansion in the form defined earlier. This issue can however in principle  be circumvented as one can recover the light cone gauge fields by gauge transforming the Lorenz gauge solutions.

To do this, we construct the unitary operator $U$ which implements the transformation from Lorenz gauge to LC gauge. This operator is given by
    \begin{equation}
        U(x^+,x^-,x_\perp) = P_{x^-}\left\{ \exp\left( -ig \int_{-\infty}^{x^-} dy^-\ {A}^+(x^+,y^-,x_\perp)\right)\right\}. 
    \end{equation}
Here, ${A}=T^aA^a$ where  $T^a$ are the generators of $SU(3)$ in the adjoint representation. We have chosen to integrate in the exponent starting at $-\infty$, which corresponds to imposing the residual gauge fixing at $x^-\rightarrow \infty$.

The fields in LC gauge are then given by
\begin{equation}
A^\mu_{LC}=U^\dagger A^\mu U-\frac{i}{g} U^\dagger\partial^\mu U
\end{equation}

When all Lorenz gauge fields are weak, the operator $U$ can be expanded to linear order in the fields. The light cone gauge fields are then calculated as
    \begin{equation}\label{lceik}
        A_{LC}(x)=A^i(x)-\partial^i\int_{-\infty}^{x^-}d y^-A^+(x^+,y^-,x_\perp).
    \end{equation}
Using the eikonal expansion for the Lorenz gauge fields we find
    \begin{eqnarray}\label{sollcg}
        A^i_{LC}&=&-\theta(x^-)\partial^ia^+_{00}
+\xi\left[\delta(x^-)a^i_{00}-x^+\theta(x^-)\partial^ia^+_{10}-\delta(x^-)\partial^ia_{01}^+\right] \\
&=& \theta(x^-)\frac{\partial^i}{\partial_\perp^2}j^+_{00}+\xi\left[\theta(x^-)x^+\frac{\partial^i}{\partial^2_\perp}j^+_{10}-\delta(x^-)\frac{1}{\partial^2_\perp}j^i_{00}+\delta(x^-)\frac{\partial^i}{\partial_\perp^2}\left(j^+_{01}+\frac{2}{\partial_\perp^2}j^+_{10} \right)\right]\nonumber
 \end{eqnarray}
Note that in the weak field limit, to leading order in the field the color currents do not change when transformed from Lorenz to Light Cone gauge. Thus $j^\mu_{\alpha\beta}$ appearing on the RHS of \eqref{sollcg} can be understood as the color currents in the Light Cone gauge.

This section has established that in the weak field limits one can apply eikonal expansion in  Lorenz or Light Cone gauges to any order. Before moving on to the question of what happens when nonlinear corrections are important, we make a detour and consider the classical fields that arise in the Born-Oppenheimer approach to quantum evolution developed in \cite{BO1}.

\section{Classical fields and the Born-Oppenheimer Renormalization group approach}

In this section we discuss a different aspect in which classical fields are important in the quantum evolution. In a recent set of papers \cite{BO1}, \cite{BO2} the authors developed a formulation of quantum evolution that encompasses both the BFKL and DGLAP kinematics. The parameter of this evolution is the frequency of the gluon field modes rather than their longitudinal momentum. The first step in this approach is to find the ground state wave function of the fast modes in the background of fixed slow modes. Subsequently one includes faster modes in the evolution, while treating more modes as the slow background. Here we are interested in the first step of this program, i.e. the wave function of fast modes.

Ref. \cite{BO1} found the  ground state wave function of a fast mode ($p$) in the background of slow modes in the form

\begin{equation}\label{wf}
|\Psi_F\rangle= e^{G(p^+, p_\perp)}|0\rangle
\end{equation}
with
\begin{equation}\label{ca}
\begin{split}
G(p^+, \v p_\perp)= A^{\dagger a }_{i}(p^+,  \v p_\perp)C_i^a(p^+,  \v p_\perp)-  A_{i}^a(p^+,  \v p_\perp)C_i^{\dagger a}(p^+,  \v p_\perp)
\end{split}
\end{equation}
where
\begin{equation}\label{c}
\begin{split}
C_i^{a}(p^+,   \v p_\perp)
=&-
 g \int_{p^->k^-, p^->(k-p)^-} 
  \frac{2p^+(k^+-p^+)}{k^+}\, \\
&\times\left\{\left[\delta_{kl}\delta_{ji}\left(\frac{2k^+}{p^+}-1\right)+\epsilon_{lk}\epsilon_{ji}\right]+\left[\delta_{ki}\delta_{jl}\left(\frac{2k^+}{k^+-p^+}-1\right)+\epsilon_{ik}\epsilon_{jl}\right]\right\} \\
&\times \frac{ \v p_{\perp j}}{ \v p_\perp^2} a^{\dagger b}_{l}(k^+-p^+, \v k_\perp- \v p_\perp)f^{abc}  a_{k}^c(k^+, \v k_\perp)\\
=&-igf^{abc}\int_{p^->all}\frac{2p^+(k^+-p^+)}{\bm{p}_\perp^2k^+}\bigg[\frac{2k^+}{p^+}p_\perp^ia_b^{j \dagger}(k-p)a^j_c(k)-\\
&    -2p_\perp^ja^{i \dagger}_b(k-p)a^j_c(k)+\frac{2k^+}{k^+-p^+}p_\perp^ja^{j \dagger}_b(k-p)a^i_c(k)\bigg],
\end{split}
\end{equation}
where $p^-\equiv p_\perp^2/p^+$, $p\equiv (p^+,p_\perp)$ and the integration measure is defined as $\int_k\equiv \int \frac{dk^+}{2\pi}\frac{d^2k_\perp}{(2\pi)^2}$.

Here $A^a_i(p)$ is the dynamical gluon field for which the ground state wave function is being sought, meanwhile $a^a_i(k)$ and $a^a_i(p-k)$ are the slow fields with $k^-\ll p^-$ and $(k-p)^-\ll p^-$.  Equation \eqref{c} holds when the slow fields are also hierarchical, i.e. $k^-<(k-p)^-$.

In the eikonal regime, where one assumes $p^+\ll k^+$, the function $C_i^a$ reduces to the static "classical field" produced by the "valence" modes that appears in the leading eikonal approximation and the JIMWLK wave function\cite{misha}, \cite{misha1}.

The wave function \eqref{wf} is that of a coherent state of the fast mode. This suggests that also beyond the strict eikonal limit it can be interpreted as producing a "classical field" $C^a_i$  of the fast gluon mode.
Indeed, it is easy to see that the expectation value of the fast field in the  state \eqref{wf} is
\begin{equation}\label{exv}
\langle\Psi_F|A^a_i(p)|\Psi_F\rangle= \frac{1}{2p^+}C_i^a(p)
\end{equation}

 In this section we  show that this field indeed has an interpretation as a solution of classical EOM in the presence of sources generated by the slow field $a$, and analyze their eikonal content.

\subsection{Born-Oppenheimer vacuum and classical fields}
The calculations of \cite{BO1} are performed assuming weak fields, and we will therefore work in the weak field limit in this section. 

Consider classical equations of motions of the Yang-Mills theory in the presence of external color currents $J_\mu^a$. The currents are comprised from the slow fields $a_\mu$:
\begin{equation}
    J^\mu=gf^{abc}\partial_\nu(a^\nu_b a^\mu_c)-gf^{abc}(\partial^\nu a^\mu_b-\partial^\mu a^\nu_b)a^c_\nu
\end{equation}
 or
\begin{equation}\label{bocur}
    \begin{aligned}
        &J^+=-gf^{abc}(\partial^+a^i_b)a^i_c\\
    &J^i=-gf^{abc}\left[2(\partial^+a^i_b)a^-_c-2(\partial^ja^i_b)a^j_c-(\partial^+ a^-_b)a^i_c+(\partial^ia^j_b)a^j_c+(\partial^ja^j_b)a^i_c\right].
    \end{aligned}
\end{equation}

In the light cone gauge $A^+=0$, the  linearized equations of motion for the transverse components of $A^a$, assuming the sources are static ($x^+$ independent),  reduce to 
\begin{equation}\label{cs}
    \partial^2_\perp A^i_a(x^-,\v x)=\frac{\partial^i}{\partial^+}J_a^+(x^-,\v x)-J_a^i(x^-, \v x)
\end{equation}
It is a straightforward matter to see that in momentum space
\begin{equation}\label{j}
\begin{split}
&\frac{\partial^i}{\partial^+}J^+(p^+,\v p)-J^i(p^+, \v p)=-\int_{k}\bar C_i^a(p,k);\\
&\bar C_i^a(p,k)\equiv igf^{abc}
\bigg[\frac{2k^+}{p^+}p^ia_b^{j \dagger}(k-p)a^j_c(k)-\\
&    -2p^ja^{i \dagger}_b(k-p)a^j_c(k)+\frac{2k^+}{k^+-p^+}p^ja^{j \dagger}_b(k-p)a^i_c(k)\bigg],
\end{split}
\end{equation}

Comparing eqs\eqref{cs} and \eqref{j} with \eqref{exv}, we see that they are very similar, but not identical. The difference is in the factor $(k^+-p^+)/k^+$. This factor is unity in the eikonal limit, but is a number of order one in the DGLAP regime, where the wave function \eqref{wf} also applies.  Since this particular factor in \eqref{exv}, \eqref{c} arises due to an energy denominator in which  the frequency of the slower fields was not completely neglected relative to $p^-$, this suggests that we should not treat the currents as completely static, but include their slow variation in time.

There is a simple analogy that we follow here. Consider a trivial case of a (classical) driven harmonic oscillator. 
    \begin{equation}
        H=\left(\omega_0-i\beta\right)\left(a^\dagger a+\frac{1}{2} \right)-\alpha\left(a^\dagger e^{-i\omega t} + ae^{i\omega t}\right),
    \end{equation}
    with $\omega_0\gg\omega$.
    We have added here a small imaginary part to the frequency, that plays the role of friction, and will eventually be taken to zero, $\beta\rightarrow 0$. It is needed to imitate the infinitesimal tilt of the contour into the imaginary time plane needed for the validity of Low's theorem, which was used to derive \eqref{wf} in \cite{BO1}.
    
    The time dependence of the 
"annihilation operator"  $a$ is given by
    \begin{equation}\label{a(t) forced harmonic oscillator}
        \frac{da}{dt} = -(i\omega_0+\beta)a+i\alpha e^{-i\omega t}.
    \end{equation}
    The general solution is of course a sum of a solution to the homogeneous equation, which is an oscillation with frequency $\omega_0$, and a particular solution with the frequency of the driving force $\omega$.
At long times the (infinitesimal) friction  (that is, $t\gg\beta^{-1}$ and $\beta\rightarrow 0^+$) kills the homogeneous solution, with the result 
    \begin{equation}
        a(t)\rightarrow_{t\rightarrow \infty}\frac{\alpha}{\omega_0-\omega}e^{-i\omega t}. 
    \end{equation}
    This behavior is quasi-static, since the variation is on a time scale much greater than the natural time scale of the oscillation of $a$. At this point we can neglect the time dependence, with the result 
    \begin{equation}\label{aqs}
    a=\frac{\alpha}{\omega_0-\omega}
    \end{equation}
    If we were to neglect the time dependence of the source from the outset, we would find the static solution $a=\alpha/\omega_0$. This is the analog of \eqref{cs}, while the quasi static solution \eqref{aqs} is the analog of \eqref{exv}.

  Let us see explicitly how this arises  in our problem. We take the slow fields to be solutions of the free equation of motion 
    \begin{equation}\label{Background field time dependence}
        a^\dagger(k) \rightarrow a^\dagger(k,x^+) = a^\dagger(k)e^{ik^-x^+}; \quad a(k)\rightarrow a(k,x^+)=a(k)e^{-ik^-x^+}
    \end{equation}
The classical equation of motion for $A_i^a$ is then
    \begin{equation}
        \frac{d}{dx^+}A^a_i(p,x^+) = -ip^-A^a_i(p,x^+)+\frac{i}{2p^+}\int_k\bar{C}^a_{i}(p,k)e^{-i[k^--(k-p)^-]x^+}.
    \end{equation}
    The particular solution that dominates at long time is
        \begin{equation}\label{tbo}
        A^a_i(p,x^+) = \frac{1}{2p^+}\int_k\frac{1}{(p^--k^-+(k-p)^-)}\bar{C}^a_{i}(p,k)e^{-i[k^--(k-p)^-]x^+}. 
    \end{equation}
    For $p^-\gg(k-p)^->k^-$
we have
    \begin{equation}
     p^- + (k-p)^- - k^- \approx \frac{\bm{p}^2k^+}{2p^+(k^+-p^+)},
    \end{equation}
and our solution becomes
    \begin{equation}\label{tdsol}
        A^a_i(p,x^+) = -\frac{1}{2p^+}\int_k \frac{2p^+(k^+-p^+)}{\bm{p}^2k^+}\bar C^a_{i}(p,k)e^{i(k-p)^-x^+}
    \end{equation}
Now neglecting the oscillating factor we recover exactly \eqref{exv}.  

Thus we see that indeed the ground state wave function of the fast fields creates the quasi static solution of classical equations of motion in the background of slow fields.

\subsection{Eikonally expanding Born-Oppenheimer fields}
In view of our discussion of the eikonal expansion, it is interesting to ask what terms are contained in the eikonal expansion of the classical solution 
in the Born-Oppenheimer approximation. We consider only expansion to $O(\xi)$. To this order the expansion of a general classical solution in the light cone gauge is given in \eqref{sollcg}.

On the other hand we can directly eikonally expand  the classical field \eqref{tbo}. This is easily done in momentum space, remembering that expansion in derivatives of $\delta(x^-)$ maps directly into expansion in powers of $p^+$ in momentum space. This expansion starts with the term $1/p^+$, as it corresponds to the $\theta(x^-)$ terms in \eqref{sollcg}.
Also, as before we need to expand in powers of $x^+$.
Using
\begin{equation}
    \begin{aligned}
        &A^{\dagger}(k^+-p^+,k-p)=A^\dagger(k^+,k-p)-p^+\frac{\partial A^{\dagger}}{\partial k^+}(k^+,k-p)+\cdot \cdot \cdot\\
        &\frac{2k^+}{k^+-p^+} \approx2\left(1+\frac{p^+}{k^+}\right)\\
        &e^{-i[k^--(k-p)^-]x^+}\approx1-ix^+[k^--(k-p)^-]\approx 1-ix^+\left(\frac{2kp-p^2}{2k^+}\right)
    \end{aligned}
\end{equation}
we find for the BO solution
\begin{equation}\label{boeik}
A^a_i(\v p, x^-,x^+)=\theta(x^-)T_1(\v p)+\delta(x^-)T_2(\v p)+x^+\theta(x^-)T_3(\v p)
\end{equation}

with
\begin{equation}\label{boeik1}
    \begin{aligned}
        &T_1(\v p)=igf^{abc}\int_k\frac{2k^+}{\bm{p}^2p^+}p^ia^{j \dagger}_b(k^+,\bm{k}-\bm{p})a^j_c(k^+,\bm{k})\\
        &T_2(\v p)=igf^{abc}\int_k\Big[\frac{-2p^i}{\bm{p}^2}a^{j \dagger}_b(k^+,\bm{k}-\bm{p})a^j_c(k^+,\bm{k})-\frac{2k^+}{\bm{p}^2}p^i\left(\frac{\partial}{\partial k^+}a^{j \dagger}_b(k^+,\bm{k}-\bm{p})\right)a^j_c(k^+,\bm{k})-\\
        &\ \ \ \ \ \ \ \ -\frac{2p^j}{\bm{p}^2}a^{i \dagger}_b(k^+,\bm{k}-\bm{p})a^j_c(k^+,\bm{k})+\frac{2p^j}{\bm{p}^2}a^{j \dagger}_b(k^+,\bm{k}-\bm{p})a^i_c(k^+,\bm{k})\Big]\\
        &= igf^{abc}\frac{1}{\bm{p}^2}\int_k \Big[2k^+p^ia^{j \dagger}_b(k^+,\bm{k}-\bm{p})\frac{\partial}{\partial k^+}a^j_c(k^+,\bm{k})-2p^ja^{i \dagger}_b(k^+,\bm{k}-\bm{p})a^j_c(k^+,\bm{k})
        +2p^ja^{j \dagger}_b(k^+,\bm{k}-\bm{p})a^i_c(k^+,\bm{k})\Big]\\
        &T_3(\v p)=gf^{abc}\int_k\frac{2\bm{k}\cdot\bm{p}-\bm{p}^2}{\bm{p}^2p^+}p^ia^{j \dagger}_b(k^+,\bm{k}-\bm{p})a^j_c(k^+,\bm{k})
    \end{aligned}
\end{equation}

It is now straightforward to compare this with the eikonal expansion of general vector potentials \eqref{sollcg}.  Using the explicit expressions of the color currents in terms of the slow fields, \eqref{bocur}, we find 
\begin{eqnarray}
T_1(\v p)&=&-\frac{p^i}{p^2}\int d^2\v xe^{-i\v p\v x}\int_{-\infty}^{\infty} dx^-J^+(0,x^-,\v x)=-\frac{p^i}{p^2}\int d^2\v xe^{-i\v p\v x}J^+_{00}(\v x)\\
T_2(\v p)&=&-\frac{1}{\v p^2}\int d^2\v xe^{-i\v p\v x}J^i_{00}(\v x)+\frac{\v p^i}{\v p^2}\left(\int d^2\v xe^{-i\v p\v x}J^+_{01}(\v x)+\frac{2}{\partial_\perp^2}\int d^2\v xe^{-i\v p\v x}J^+_{10}(\v x) \right) \nonumber\\
T_3(\v p)&=& \frac{\v p^i}{\v p^2} \int d^2\v xe^{-i\v p\v x}J^+_{10}(\v x)\nonumber
\end{eqnarray}
Noting that
\begin{equation}
        \begin{aligned}
            J^+_{00} & = \xi^0 j^+_{00} \\
            J^+_{10} & = \xi j^+_{10} \\
            J^+_{01} & = \xi j^+_{01}\\
            \vdots
        \end{aligned}
    \end{equation}
we see that the eikonal expansion of the BO classical fields reproduces all terms in the general eikonal expansion. 

Interestingly, we have to keep the time dependence in the classical solution in order to recover all terms. This is obvious for the term proportional to $x^+$, as without nontrivial time dependence such a term cannot appear in principle. However the time dependence is also crucial to recover $T_2$ correctly. In particular, the first term in $T_2$ in \eqref{boeik1} arises from expanding the prefactor in \eqref{tdsol} to subleading order. As discussed in the previous subsection, this term would not be present if we did not keep the time dependence in the quasi-static time dependence of the source in the Yang-Mills equations.

\section{Eikonal expansion for strong fields }

So far we have considered the weak field regime, where nonlinear terms in the equations of motion are negligible. The eikonal expansion in this regime is well defined and can be pursued to any order in $\xi$. However for the applications, in which  we are often interested the fields are not weak, but parametrically large $A\sim 1/g$. This is the case, for example in discussions of saturation.
\subsection{Nonlinear corrections mix eikonal orders}

Let us therefore now consider the situation where the color fields are large.    The solution of the EOM in Lorenz gauge formally is the same as \eqref{sol2}. However the quadratic term in the equation of motion generates a term which is not of the same form as any of the terms in the eikonal expansion of the current, and thus in addition we find at the first subeikonal order a constraint
    \begin{equation}\label{constr}
        \xi^1: 2ig[\delta(x^-)]^2\left(\big[a^+_{00},a^+_{10}\big] +\big[ \partial^ja^+_{00},a^j_{00}\big]\right) = 0.\ ,
    \end{equation}
    which using the Lorenz gauge condition becomes
    \begin{equation}\label{constr1}
     2ig[\delta(x^-)]^2\partial^j\big[a^+_{00},a^j_{00}\big]= 0
     \end{equation}

This looks puzzling since a solution of \eqref{sol2} does not automatically satisfy \eqref{constr1}.
It thus looks that the eikonal expansion is inconsistent with nonlinearities in the EOM.

Let us try to understand the reason for the appearance of this constraint. Formally it arises as the coefficient of the $\delta(x^-)$ term in the EOM at order $\xi$. However, the coefficient of this term is divergent as it involves $\delta(0)$. To define this coefficient one would have to regulate the $\delta$-function. The only natural scale for such a regulator would be the longitudinal size of the current distribution, which is of order $\xi$. Thus for a regularized $\delta$-function one expects $\delta(0)\sim \xi^{-1}$. If this is the case, the term \eqref{constr} is in fact $O(1)$ rather than $O(\xi)$. Thus it should not be considered as a constraint, but as an addition to the $O(1)$ EOM that determines $a^+_{00}$.

To understand this better let us calculate components of $F^{\mu\nu}$ in terms of $a^\mu_{mn}$.
To $O(\xi)$ we find
    \begin{equation}
        \begin{aligned}
            &F^{i-}  = 0\\
            &F^{-+}  = \xi\delta(x^-)a^+_{10} \\
            &F^{ij}  = \xi\delta(x^-)\left( \partial^ia^j_{00}-\partial^ja^i_{00}\right)\ ,
        \end{aligned}
    \end{equation}
and importantly
  \begin{equation}\label{fi+}
        \begin{split}
            F^{i+} = \delta(x^-)\partial^ia^+_{00}+\xi\Big( x^+\delta(x^-)\partial^ia^+_{10}+ \delta'(x^-)\partial^ia^+_{01} -\delta'(x^-)a^i_{00}\Big)+ig\xi [\delta(x^-)]^2\left[a^i_{00},a^+_{00} \right]
        \end{split}
    \end{equation} 
We recognize the last term in \eqref{fi+} as the problematic term \eqref{constr1}. The issue now is clear. Suppose we were to expand directly the field strength $F^{i+}$ in $\xi$ without assuming that the vector potentials themselves have an eikonal expansion. We would then have at leading order
\begin{equation}
F^{i+}=\delta(x^-)f_{00}^{i+}(x_\perp)+O(\xi)
\end{equation}
with
\begin{equation}\label{f00}
f_{00}^{i+}(x_\perp)=\int dx^-\Big\{\partial_ia^+(0,x^-,x_\perp)+ig[a^i(0,x^-,x_\perp),a^+(0,x^-,x_\perp)]\Big\}
\end{equation}
The second term in this equation is in general not negligible and  contributes to $f_{00}^{i+}$. On the other hand if we decompose eikonally the vector potential and calculate $F^{i+}$ using their expanded form, we obtain \eqref{fi+} where the last term is formally of order $\xi$. As we have noted above, once $\delta(0)$ is regulated on the distance scale $\xi$ this term in fact becomes $O(1)$ and thus contributes to the leading eikonal order.

Higher order eikonal terms $a^\mu_{ij}$ 
 are also problematic in the calculation of   $F^{i+}$ . 
For example, keeping the terms which are formally of order $\xi^2$, we get an additional contribution
\begin{equation}
+ig\xi^2\delta(x^-)\delta'(x^-)\Big([a^+_{01}(x_\perp),a^i_{00}(x_\perp)]+[a^+_{00}(x_\perp),a^i_{01}(x_\perp)]\Big)
\end{equation}
This again needs regularization in order to be properly defined, and with the natural regulator $\delta(0)\rightarrow 1/\xi$ becomes
\begin{equation}
\sim ig\xi\delta'(x^-)\Big([a^+_{01}(x_\perp),a^i_{00}(x_\perp)]+[a^+_{00}(x_\perp),a^i_{01}(x_\perp)]\Big)
\end{equation}
with a regularization dependent prefactor. This is therefore a finite correction to the $O(\xi)$ term in \eqref{fi+} even though formally we could have expected it to be $O(\xi^2)$. In fact there is an infinite number of such corrections that contribute upon regularization to any given eikonal order, and thus one needs to resum an infinite number of terms at any eikonal order.
The same applies to other components of $F^{\mu\nu}$: formally higher order eikonal corrections turn into leading order contributions upon regularization.


The consequence of this is that the eikonal expansion does not operate as a well defined perturbative expansion. One does not have an autonomous equation for the "leading" term, which subsequently would allow one to determine higher order terms perturbatively. Instead subeikonal components couple to the leading eikonal component already at the leading order via the nonlinear terms in the classical equations of motion.

Thus it looks like the eikonal expansion of the vector potential is of very little use for the calculation of $F^{\mu\nu}$, as well as for expanding the EOM.

Things are not so bad if the nonlinear terms in physical quantities, like $F^{\mu\nu}$ are suppressed by an external factor independent of $\xi$. Such a factor could be for example the QCD coupling constant. Thus if the subeikonal components of spatial fields $a^i$ are suppressed by a power of $\alpha_s$, all the nonlinear contributions are suppressed as well. One can then neglect the last term in \eqref{f00} relative to the other terms, and correspondingly the last term in \eqref{fi+} relative to the first term. 
This state of things has been noted in \cite{ming}, which considered the regime where the subeikonal components of vector potential are suppressed by a power of the coupling constant. In this regime the expansion makes sense, and can be applied to solve the EOM.

Unfortunately, the assumption that all subeikonal terms are suppressed by the QCD coupling constant is unnatural. One could argue classically that the time ($x^+$) dependence of the vector potential may be suppressed, and thus $a^\mu_{n\alpha}$ for $n\ge 1$ are small.  In the absence of interaction any static distribution of sources is stable, and it is therefore possible to have a static $J^+$ with vanishing $J^i$ such that all $j^\mu_{n\alpha}$ (and therefore $a^\mu_{n\alpha}$) components for $n\ge 1$ vanish.  Turning on interactions will necessarily create currents $J^i$, but they will be suppressed by $\alpha_s$ relative to the static component of $J^+$. Due to the conservation equation of course, one cannot have large $j^+_{n 0}$ for $n\ge 1$ without also having large $j^i$, thus if $J^i$ is suppressed, so are $J^+_{n0}$. However
there is no reason, for example for the second moment of the distribution in $x^-$ (i.e. $j^+_{01}$) to be parametrically smaller than the first moment ($j^+_{00}$) if we assume an arbitrary smooth distribution of charges and currents in the rest frame. Thus generically one expects different $x^-$ moments of a given component of a current to be of the same order in $\alpha_s$.

One is then left with a problem. It looks like for strong fields, different eikonal components of a given field mix, and even at leading eikonal order one does not have a closed equation that would determine $a^+_{00}$. But the standard approach in the modern literature to low $x$ physics \cite{mv} starts with the assumption that $a^+_{00}$ is determined directly by $j^+_{00}$. To be sure one simply puts all other eikonal components to zero when determining $j^+_{00}$ under a tacit assumption that even if included, they do not change the solution at leading order in $\xi$. Our analysis above shows that this is not a valid procedure for strong fields: nonvanishing $a^\mu_{0n}$ with $n\ge 1$ do not decouple from $a^+_{00}$ even at leading eikonal order. Worse than that, the coupling between modes cannot be determined without prescribing the regulator for $\delta(x^-=0)$.

So, we are faced with the following question: is the standard treatment of the leading eikonal contribution inconsistent? The answer is that one regains consistency if the eikonal expansion is reinterpreted somewhat differently. The rest of this section is devoted to the discussion of this point.

\subsection{Eikonal expansion mark II: "integrating out" high momentum modes.}
As we have seen above, the straightforward eikonal expansion of vector potentials is problematic, or better to say of very little practical use when nonlinear corrections to equations of motion are non negligible.  To understand this a little better let us first take a look at the eikonal expansion in momentum space. 

\subsubsection{Going to momentum space}
In the rest of the paper we do not explicitly indicate the part of eikonal expansion that arises from expanding in powers of $x^+$. These terms are harmless since nonlinearities in the fields and Lagrangian/Hamiltonian densities are local in $x^+$. We concentrate exclusively on expansion in derivatives $\delta^{(n)}(x^-)$. We also do not explicitly write the transverse coordinate, color and Lorentz indices for simplicity. Thus our notation from now on is such that
\begin{equation}
    a_n\equiv \sum_ma^a_{inm}(x_\perp)(\xi x^+)^m
\end{equation}

In momentum space, the higher derivatives of $\delta(x^-)$ of course correspond to higher powers of momentum. Thus for an arbitrary function $A(x^-)$ (which could be any component of vector potential or color current) in Fourier space the eikonal expansion reads
\begin{equation}\label{peik}
    A(p^+)=\xi^{\alpha}\Big[a_0-i\xi p^+a_1+(i\xi p^+)^2a_2+\cdots\Big]
\end{equation}
where $a_i$ are numbers (functions of $x_\perp$ and $x^+$) of order unity, in units of the strong interaction scale $\Lambda_{QCD}$, and the power $\alpha$ depends on the field one is expanding.
This Taylor expansion in momentum space is good as long as $p^+\ll \Lambda_{QCD}/\xi$. 
For larger $p^+$ the Taylor expansion is not useful, in the sense that all terms become of the same order and thus all must be resummed.

From this point of view the problem we encountered above stems from the fact that a product of two functions in coordinate space becomes a convolution in momentum space. Calculating a convolution of any two terms in the eikonal expansion \eqref{peik} we have
\begin{equation}
    \xi^{2\alpha+l+m}a_la_m\int dp^+ (p^+)^{l+m}\sim \xi^{2\alpha+l+m}a_la_m \mu^{l+m+1}
\end{equation}
where $\mu$ is the cutoff on the momentum necessary to regulate the integral. Physically this cutoff is of order $\Lambda_{QCD}/\xi$, provided typical momenta in the hadron rest frame are of order $\Lambda_{QCD}$. Thus we again see that the convolution of two terms in eikonal expansion is of a lower eikonal order than we would naively expect.

However we should remember that our interest in eikonal expansion is mainly because we hope to use it as a convenient tool for low $x$ physics. In this setting we are only interested in the low $p^+$ components of the various fields in the problem. Thus we do not have to eikonally expand the field $A$ for all momenta, but only for some low momenta $p^+<\Lambda$. For low $x$ physics the scale $\Lambda$ is naturally  
\begin{equation}\label{scaling} \Lambda= x \frac{\mu}{\xi}; \ \ \ \ \ \mu\sim \Lambda_{QCD}
\end{equation}
with $x\ll 1$.

Let us therefore define
\begin{eqnarray}\label{lowma}
 A_L(p^+,x^+,x)&=&A(p^+,x^+,x)\theta\left(1-\frac{|p^+|}{\Lambda}\right)
 \\
 &\approx&
 \xi^{\alpha}\Big[a_0(\xi x^+,x)+i\xi p^+a_1(\xi x^+,x)+(i\xi p^+)^2a_2(\xi x^+,x)+...\Big]\theta\left(1-\frac{|p^+|}{\Lambda}\right)\nonumber
 \end{eqnarray}
 where we have assumed that at momenta $p^+<\Lambda$ the eikonal expansion of the field $A$ makes sense.


The field $A(p^+)$ then can be written as
\begin{equation}\label{dec}
    A(p^+)=A_L(p^+)+A_H(p^+)
\end{equation}
with $A_L$ given by \eqref{lowma} and 
\begin{equation}A_H(p^+)=A(p^+)\left[1-\theta\left(1-\frac{p^+}{\Lambda}\right)\right]
\end{equation}
Obviously $A_L$ has support on low longitudinal momentum modes while $A_H$ has support on high momentum modes.

The eikonal expansion also involves expanding the fields in a Taylor series in time $x^+$. Fortunately, since the action density and EOM are local in $x^+$, this expansion does not present any problems similar to those associated with expansion in powers of longitudinal momentum.
We will therefore only split the fields in longitudinal momentum, while keeping the full dependence on $x^+$ in the functions $a_\alpha$ in 
\eqref{lowma}. One of course has to remember that further expansion in $\xi x^+$ is required to define complete eikonal expansion.


\subsubsection{EOM for low longitudinal momentum modes}
We can now write the EOM for $A_L$. Those follow from the full equations of motion in momentum space upon decomposing the fields according to \eqref{dec}. The result qualitatively is easy to understand. The linear part of $D^\mu F_{\mu\nu}$ at low longitudinal momentum contains only $A_L$ by definition. The nonlinear part  contains two types of terms. One type involves only $A_L$, and the other one only $A_H$. 

Usually in similar situations one neglects terms which are linear in  $A_H$ (whether multiplied by $A_L$ or not). Those terms would have high net longitudinal momentum and do not contribute to the equation of motion for $A_L$  by momentum conservation. This last statement is not strictly true, as there is some overlap in momentum between $A_H$ and, for example, $A_L^2$ in the region close to the separation boundary. The effect of these terms is nevertheless suppressed relative to nonlinear terms in $A_H$, as in the latter case the integration region that contributes in the equation is not limited to that close to the separation boundary. We will therefore not consider these terms in the following.

In this way the main effect of the high momentum modes is to "renormalize" the components of the color current which sources the low momentum modes. Additionally the high momentum modes generate contributions that "renormalize" the kinetic term of $A_L$.

Rather than writing the equations of motion, we can substitute the decomposition  \eqref{dec} directly into the Yang-Mills Lagrangian to obtain the interaction between various eikonal field components. This approach is somewhat more intuitive as it follows similar logic to derivation of an effective low energy Lagrangian in various physics contexts. We will therefore follow this approach below.

\subsubsection{The "self interaction" terms.}
We first consider the terms in the action that involve the low momentum fields only.
We will calculate the action to the lowest nontrivial order in $\xi$ that involves self interaction between the low momentum modes, which is $O(\xi^3)$. 
Calculation of higher order terms is straightforward if desired, but rather tedious. At any rate, the terms we keep  are sufficient to understand the general pattern, as they include interactions between the leading eikonal and subeikonal moments.

We start by calculating the components of the field strength tensor  $F^{\mu\nu}$
\begin{eqnarray}
    F^{aij}(p)&=&\partial^iA^{aj}(p)-\partial^jA^{ai}(p)-gf^{abc}\int dkA^{bi}(p-k)A^{cj}(k)\\
    &=&\xi [\partial^ia_{0}^{aj}-\partial^ja_{0}^{ai}]\theta(\Lambda-|p|)\nonumber\\
&&+\xi^2\left[ip[\partial^ia^{aj}_{1}-\partial^ja^{ai}_{1}]\theta(\Lambda-|p|)-g(2\Lambda-|p|)f^{abc}a^{bi}_{0}a^{cj}_{0}\theta(2\Lambda-|p|)\right]\nonumber\\
    F^{a+i}(p)&=&-\partial^ia^{a+}_{0}\theta(\Lambda-|p|)\\
&&+\xi\left[\left(- ip\partial^ia^{a+}_{1}+ipa^{ai}_{0}\right)\theta(\Lambda-|p|)-gf^{abc} (2\Lambda-|p|)a^{b+}_{0}a^{ci}_{0}\theta(2\Lambda-|p|)\right]\nonumber\\
&&+\xi^2\left[   \left(p^2\partial^ia^{a+}_{2} -p^2a^{ai}_{1}\right)\theta(\Lambda-p)-i\frac{g}{2}f^{abc}p(2\Lambda-|p|)[a^{b+}_{0}a^{ci}_{1}+a^{b+}_{1}a^{ci}_{0}]\theta(2\Lambda-|p|)\right]\nonumber\\
    F^{a-i}(p)&=&\xi^2\left[\frac{1}{\xi}\partial^-a^{ai}_{0}-\partial^ia^{a-}_{0}\right]\theta(\Lambda-|p|)\nonumber\\
    &&-\xi^3g(2\Lambda-p)f^{abc}a^-_{0}a^i_{0}\theta(2\Lambda-p)\\
    F^{a+-}(p)&=&- \xi[\frac{1}{\xi}\partial^-a^{a+}_{0}]\theta(\Lambda-|p|)+\xi^2\left[ip a^{a-}_{0}\theta(\Lambda-|p|) -gf^{abc}(2\Lambda-|p|)\ a^{b+}_{0}a^{c-}_{0}\theta(2\Lambda-|p|)\right]\nonumber
\end{eqnarray}
Note that every derivative $\partial^-$ brings a factor of $\xi$ in the eikonal expansion, hence the terms are organized in the above equation as they are.

To order $\xi^3$ we have
\begin{eqnarray}
\int_p F^{a+i}(p)F^a_{+i}(-p)&=&2\xi^2\Lambda \partial^ia^{a+}_{0}[\frac{1}{\xi}\partial^-a^{ai}_{0}-\partial^ia^{a-}_{0}]\\
&&+3g\xi^3\Lambda^2f^{abc}\Big\{[\frac{1}{\xi}\partial^-a^{ai}_{0}-\partial^ia^{a-}_{0}]a^{b+}_{0}a^{ci}_{0}-\partial^ia^{a+}_{0}a^{-b}_{0}a^{ci}_{0}\Big\}\\
\int_pF^{aij}(p)F^{aij}(-p)&=&\xi^2 2\Lambda[\partial^ia_{0}^{aj}-\partial^ja_{0}^{ai}]^2\nonumber\\
&&-\xi^36g\Lambda^2f^{abc}[\partial^ia^{aj}_{0}-\partial^ia^{aj}_{0}]a^{bi}_{0}a^{cj}_{0}\nonumber\\
\int_p F^{+-}(p)F^{+-}(-p)&=&\xi^2 2\Lambda \left[\frac{1}{\xi^2}\partial^-a^{a+}_{0}\partial^-a^{a+}_{0}\right]\nonumber\\
&&+\xi^3 6g\Lambda^2f^{abc}\left[\frac{1}{\xi}\partial^-a^{a+}_{0}\right]a^{b+}_{0}a^{c-}_{0}\nonumber
\end{eqnarray}
The action is given by
\begin{equation}
    S=-\frac{1}{4}\int d^4xF^{\mu\nu}F_{\mu\nu}=\frac{1}{4}\int d^4x\Big[4F^{+i}F^{-i}+2F^{+-}F^{+-}-F^{ij}F^{ij}\Bigg]
\end{equation}

The part of the action which involves only low momentum modes to order $\xi^3$ is given by
\begin{eqnarray}\label{self}
    S_{L}&=&\xi^2\Lambda\int d^3x\Big[-2\partial^ia^{a+}_{0}[\frac{1}{\xi}\partial^-a^{ai}_{0}-\partial^ia^{a-}_{0}]+\frac{1}{\xi^2}\partial^-a^{a+}_{0}\partial^-a^{a+}_{0}-\frac{1}{2}[\partial^ia_{0}^{aj}-\partial^ja_{0}^{ai}]^2\Big]\\
    &+&3g\xi^3\Lambda^2f^{abc}\Big[- [\frac{1}{\xi}\partial^-a^{ai}_{0}-\partial^ia^{a-}_{0}]a^{b+}_{0}a^{ci}_{0}+\partial^ia^{a+}_{0}a^{-b}_{0}a^{ci}_{0}+\frac{1}{\xi}\partial^-a^{a+}_{0}a^{b+}_{0}a^{c-}_{0}+\partial^ia^{aj}_{0}a^{bi}_{0}a^{cj}_{0}\Big]\nonumber\\
    &=&x\xi\mu\Big[-2\partial^ia^{a+}_{0}[\frac{1}{\xi}\partial^-a^{ai}_{0}-\partial^ia^{a-}_{0}]+\frac{1}{\xi^2}\partial^-a^{a+}_{0}\partial^-a^{a+}_{0}-\frac{1}{2}[\partial^ia_{0}^{aj}-\partial^ja_{0}^{ai}]^2\Big]\nonumber\\
    &+&3gx^2\xi\mu^2 f^{abc}\Big[- [\frac{1}{\xi}\partial^-a^{ai}_{0}-\partial^ia^{a-}_{0}]a^{b+}_{0}a^{ci}_{0}+\partial^ia^{a+}_{0}a^{-b}_{0}a^{ci}_{0}+\frac{1}{\xi}\partial^-a^{a+}_{0}a^{b+}_{0}a^{c-}_{0}+\partial^ia^{aj}_{0}a^{bi}_{0}a^{cj}_{0}\Big]\nonumber
\end{eqnarray}

\subsubsection{The interaction with the hard field.}
Next we calculate the interaction with the hard fields.
There are two types of such terms: those responsible for the renormalization of the color current, and those for the renormalization of the soft field kinetic term.

The terms linear in the soft field are easy to write. We use the simple identity 
\begin{equation}\delta F^{a\mu\nu}/\delta A^{b\lambda}=D^{ab \nu}\delta^{\mu\lambda}-D^{ab\mu}\delta^{\lambda\nu}
\end{equation}
to write
\begin{eqnarray}\label{int1}
    S^1=-\int &&\Bigg[A_L^{a+}\big[D^{abi}F^{bi-}+D^{ab-}F^{b-+}\big]_H+A_L^{a-}\big[D^{abi}F^{bi+}+D^{ab+}F^{b+-}\big]_H\\
    &&+A_L^{ai}\big[D^{abj}F^{bij} +D^{ab+}F^{b-i}+D^{ab-}F^{b+i}\big]_H\Bigg]\nonumber
\end{eqnarray}
Substituting the eikonal expansion for the low momentum fields we have (keeping only the relevant components)
\begin{equation}
    S^1=-\xi^2 \Lambda\int d^3x\Big[a^-_{0}j^+_{0R}+a^i_{0}j^i_{0R}+a^+_{0}j^-_{0R}
    \Big]
\end{equation}
with
\begin{eqnarray}\label{jays}
    j^+_{0R}&=&\frac{1}{\Lambda}\int_{-\Lambda}^{\Lambda}dp\big[D^{abi}F^{bi+}+D^{ab+}F^{b+-}\big]_H+j^{+}_{0}\\
    j^-_{0R}&=&\frac{1}{\xi^2 \Lambda}\int_{-\Lambda}^{\Lambda}dp\big[D^{abi}F^{bi-}+D^{ab-}F^{b-+}\big]_H+j^{-}_{0}\nonumber\\
    j^i_{0R}&=&\frac{1}{\xi \Lambda}\int_{-\Lambda}^{\Lambda}dp\big[D^{abj}F^{bij} +D^{ab+}F^{b-i}+D^{ab-}F^{b+i}\big]_H+j^{i}_{0}\nonumber\
\end{eqnarray}
where $j^{\mu }_{\alpha}$ are eikonal moments of the "valence" currents which source the original equations of motion.
Note that in all of these expressions only nonlinear terms on the RHS contribute, as the high momentum field vanishes for $\Lambda>p^+>-\Lambda$. The above
 is just the statement that the low momentum modes couple to the renormalized current $J^\mu_R$ which is the sum of the original "valence" current $J^\mu$ and the current due to the high momentum fields $A^\mu_H$ arising from their contribution to nonlinear part of $D_\nu F^{\mu\nu}$.
\begin{equation}
    J_{\Lambda R}^{a\mu}=J^{a\mu}+f^{abc}\partial_\nu(A_H^{b\nu} A_H^{c\mu})-f^{abc}(\partial_\nu A_H^{b\mu}-\partial^\mu A^b_{H\nu})A_H^{c\nu}
\end{equation}
The subscript $\Lambda$ here is meant to stress that the current depends on the factorization scale $\Lambda$.

Similarly for quadratic in low momentum field terms we can use
\begin{equation}
    F^{a\mu\nu}=F^{a\mu\nu}_H+ D^{\mu ab}_HA^{b\nu}_L-D^{\nu ab}_HA^{\mu b}_L-gf^{abc}A^{\mu b}_LA^{\nu c}_L
\end{equation}
Thus to extract terms quadratic in $A_L$ we have
\begin{eqnarray}
    S^2&=&-\frac{1}{2}\Big[[D^{\mu ab}_HA^{\nu b}_L][D^{ac}_{\mu H}A^{c}_{\nu L}]-[D^{\mu ab}_HA^{\nu b}_L][D^{ ac}_{\nu H}A^{ c}_{\mu L}]-gf^{abc}A_L^{\mu b}A_L^{\nu c}F^{a}_{\mu\nu H}\Big]\\
    &=&\frac{1}{2}A^{\mu a}_L\Big[    D^{\lambda ac}_HD^{cb}_{\lambda H}\delta^{\mu\nu}-D^{\nu ac}_HD^{cb}_{\mu H}+gf^{abc}F^c_{\mu\nu H}\Big]A^b_{L\nu}\nonumber\\
    &=&gf^{abc}\Big[\partial^\mu A_L^{a\nu}A_{L\nu}^{c}A^b_{H\mu}-\partial^\mu A_L^{a\nu}A_{L\mu}^{c}A^b_{H\nu}+\partial_\mu A^a_{H\nu}A_L^{b\mu}A_{L}^{\nu c}\Big]\nonumber\\
    &-&\frac{1}{2}g^2f^{abc}f^{ade}\Big[A_L^{\nu c}A_L^{\nu e}A_{H\mu}^bA_{H\mu}^d-A_L^{\nu c}A_L^{\mu e}A_{H\mu}^bA_{H\nu}^d+A_L^{\mu b}A_L^{\nu c}A_{H\mu}^dA_{H\nu}^e\Big]\nonumber\\
    &=&gf^{abc}\Big[\partial^\mu A_L^{a\nu}A_{L\nu}^{c}A^b_{H\mu}-\partial^\mu A_L^{a\nu}A_{L\mu}^{c}A^b_{H\nu}\Big]-\frac{g}{2}A^{b\mu}_LA^{c}_{\nu L}L^{bc\nu}_\mu\nonumber
\end{eqnarray}
with
\begin{equation}
L^{bc\nu}_\mu=-2f^{abc}\partial_\mu A^{a\nu}_{H}+gf^{abd}f^{ace}\Big[A^{d\lambda}_HA^e_{\lambda H}\delta_{\mu\nu}-A^{d\nu}_HA^e_{\mu H}\Big] +gf^{abc}f^{ade}A^d_{\mu H}A^{e\nu}_H
\end{equation}
We only keep the leading eikonal order contribution to each term. In addition we do not write explicitly terms that contain eikonal moments other than the ones that appear in the "self interacting" action \eqref{self}.
Using
\begin{eqnarray}
  &&  [A^{a\mu}_LA^b_{L\nu}](p)=a^{a\mu}_{0}a^b_{\nu 0}
  (2\Lambda-|p|)\theta(2\Lambda-|p|)
  \end{eqnarray}
  we get
  \begin{eqnarray} 
  &&\partial^\mu A^{a\nu} A^b_\nu=-\partial^\mu A^{aj}A^{bj}+\partial^\mu A^{a+}A^{b-}+\partial^\mu A^{a-}A^{b+}\\
  &&\mu=i\ \ \ : \ \ \ \xi^2\Big[-\partial^ia^{aj}_{0}a^{bj}_{0}+\partial^ia^{a+}_{0}a^{b-}_{0}+\partial^ia^{a-}_{0}a^{b+}_{0}\Big](2\Lambda-|p|)\theta(2\Lambda-|p|)\\
 && \mu=- \ \ :\ \ \ \xi^3[-\frac{1}{\xi}\partial^-a^{ai}_{0}a^{bi}_{0}+\frac{1}{\xi}\partial^-a^{a+}_{0}a^{b-}_{0}](2\Lambda-|p|)\theta(2\Lambda-|p|)\\
 &&\mu=+ \ \ : \ \ \ 0
\end{eqnarray}
and
\begin{eqnarray}
   &&\partial^\mu A^{a\nu} A^b_\mu=-\partial^j A^{a\nu}A^{bj}+\partial^+ A^{a\nu}A^{b-}+\partial^- A^{a\nu}A^{b+}\\   
   &&\nu=i \ \ :\ \ \ \xi^2\Big[-\partial^ja^{ai}_{0}a^{bj}_{0}+\frac{1}{\xi}\partial^-a^{ai}_{0}a^{b+}_{0}\Big](2\Lambda-|p|)\theta(2\Lambda-|p|)\\
   &&\nu=+ \ \ : \ \ \ \xi\Big[ -\partial^ja^{a+}_{0}a^{bj}_{0}+\frac{1}{\xi}\partial^-a^{a+}_{0}a^{b+}_{0}\Big](2\Lambda-|p|)\theta(2\Lambda-|p|)\\
   &&\nu=- \ \ :\ \ \ -\xi^3\partial^ja^{a-}_{0}a^{bj}_{0}(2\Lambda-|p|)\theta(2\Lambda-|p|)\\
\end{eqnarray}
Now combining terms
\begin{eqnarray}\label{int2}
    S^2&=&-g\xi^2\Lambda\Big[ (\partial^ja^{bi}_{0}-\partial^ia^{bj}_{0})a^{cj}_{0}m_i^a+a^{bi}_{0}a^{cj}_{0}m_{i}^{bcj}\Big]\nonumber\\
    &&-g\xi^2\Lambda \Big[f^{abc}[\partial^ia^{b+}_{0}a^{c-}_{0}+\partial^ia^{b-}_{0}a^{c+}_{0}]m_i^a 
    +\frac{1}{2}a^{b+}_{0}a^{c-}_{0}m^{bc+}_+\Big]\\
    &&-g\xi^2\Lambda f^{abc}\partial^ia^{b+}_{0}a^{ci}_{0}m_+^a
   \nonumber\\
    &&-\frac{g}{2}\xi^2\Lambda a^{bi}_{0}a^{c+}_{0}m^{bc-i}
    \nonumber\\
    &&-g\xi^2 \Lambda f^{abc}\partial^ia^{b-}_{0}a^{ci}_{0}m^a_-
   \nonumber\\
     &&-\frac{g}{2}\xi^2\Lambda a^{bi}_{0}a^{c-}_{0}m^{bc+i}
     +\cdots\nonumber
\end{eqnarray}
\begin{eqnarray}\label{ams}
    m_i^a&=&-\frac{1}{\Lambda}\int_{-2\Lambda}^{2\Lambda}dp(2\Lambda-|p|)A^a_{iH}(p)\nonumber\\
    m_{i}^{bcj}&=&\frac{1}{\Lambda}\int_{-2\Lambda}^{2\Lambda}dp(2\Lambda-|p|)L^{bc j}_i(p)\nonumber\\
    m^{bc+}_+&=&\frac{1}{\Lambda}\int_{-2\Lambda}^{2\Lambda}dp(2\Lambda-|p|)L^{bc+}_{+H}(p)\nonumber\\
    m^a_+&=& -\frac{1}{\xi \Lambda}\int_{-2\Lambda}^{2\Lambda}dp(2\Lambda-|p|)A^a_{+H}(p)\nonumber\\
    m^{bc-i}&=&\frac{1}{\xi\Lambda}\int_{-2\Lambda}^{2\Lambda}dp(2\Lambda-|p|)[L^{bc -i}+L^{cb i-}](p)\nonumber\\
     m^a_-&=&-\frac{\xi}{\Lambda}\int_{-2\Lambda}^{2\Lambda}dp(2\Lambda-|p|)A^a_{-H}(p)\\
     m^{bc+i}&=&\frac{\xi}{\Lambda} \int_{-2\Lambda}^{2\Lambda}dp(2\Lambda-|p|)[L^{bc +i}+L^{cb i+}](p)\nonumber
\end{eqnarray}
In the following with a slight abuse of language we will refer to the coefficients $m$ as "masses".


\subsection{The $x\rightarrow 0$ limit}
Equations\eqref{self},\eqref{int1} and \eqref{int2} together with \eqref{jays} and \eqref{ams} constitute the action describing the dynamics of the low momentum mode. This action explicitly depends on a chosen value of $x$ that defines the separation scale $\Lambda$. Of course, the solutions of the equations of motion for $a^\mu_{\alpha}$ should not depend on the value of $x$, so that the explicit $x$-dependence is compensated by the nontrivial $x$-dependence of the renormalized currents $j^\mu_R$ and masses $m^\mu_\nu$. 
The dependence of the $j's$ and $m's$ on $x$ is determined by the equation
\begin{equation}
    \frac{dS}{dx}=\frac{\partial S}{\partial x}+\sum_{n=j,m}\frac{\partial S}{\partial n}\frac{\partial n}{\partial x}=0
\end{equation}
where the summation in the second term goes over all eikonal components of $j^\mu$ and $m^{\mu}_{\nu}$.
The scale $x$ thus plays a role very analogous to that of the factorization scale in perturbative QCD calculations. In the spirit of this analogy it is natural to view the self interaction as the "hard part" of the calculation while viewing the renormalized currents and masses as analogs of nonperturbative PDF's and similar quantities.

As is commonly the case with factorization scales, it is advantageous to make the most convenient choice which makes the "perturbative" part of the calculation as simple as possible.
In this respect we observe 
that the above  refinement of the eikonal expansion has a profound effect on the equation of motion for the eikonal moments. Now that the contribution of high momentum modes has been separated out, explicit nonlinearities in the equations of motion are much milder. 
For $\Lambda$ with scaling of \eqref{scaling}, the "mixing" of eikonal orders still occurs; however, the nonlinear terms in the EOM for the soft modes are suppressed by powers of $x\ll 1$. Thus these terms can be treated perturbatively in the expansion in powers of $x$. 
In fact it becomes obvious that it is most advantageous to choose $x=0$. With this choice all the self interactions vanish and all nontrivial effects are shifted into the renormalization of currents and masses. The choice $x=0$ is quite analogous to the choice $\mu^2=Q^2$ in DIS, which moves all large corrections to nonperturbative PDF's and leaves the hard part of the process maximally simple.

The interesting question is therefore how the currents and masses scale in the limit $x\rightarrow 0$. We can answer this question by examining \eqref{jays} and \eqref{ams}. Consider for example $j^+_{0R}$. The limit $x\rightarrow 0$ corresponds to $\Lambda\rightarrow 0$.
In this case the integration measure cancels the factor $1/\Lambda$, and we get
\begin{eqnarray}\label{j0}
 &&   j^+_{0R}\rightarrow_{x\rightarrow 0} 2\int dx^-\big[D^{abi}F^{bi+}+D^{ab+}F^{b+-}\big]_H+j^{+}_{0}\\
 &&=2\int dx^-\big[D_R^{abi}f^{bi+}+D_R^{ab+}f^{b+-}\big]_H+j^{+}_{0}
\end{eqnarray}
Here the vector potential that enters in the definition of $f^{\mu\nu}$ and $D_R$ is the solution of the full equation of motion (with $\Lambda\rightarrow 0$), and the subscript $H$ on parenthesis indicates that only nonlinear terms in this vector potential are included on the RHS.  As per our original assumption, the integral in \eqref{j0} is finite. 
Similarly
\begin{eqnarray}
&&   j^-_{0R}\rightarrow_{x\rightarrow 0} =2\int dx^-\big[D_R^{abi}f^{bi-}+D_R^{ab-}f^{b-+}\big]_H+j^{-}_{0}\\
&&   j^i_{0R}\rightarrow_{x\rightarrow 0} =2\int dx^-\big[D_R^{abj}f^{bij}+D_R^{ab+}f^{b-i}+D_R^{ab-}f^{b+i}\big]_H+j^{i}_{0}
\end{eqnarray}
are finite.
Thus in the limit $x\rightarrow 0$ the eikonal moments of the current undergo finite renormalization. 


Next is the question of masses. Here the situation is very simple. Using the same arguments as above we find that all the mass coefficients vanish in the interesting limit
\begin{equation}
    m_\gamma\propto \xi\Lambda=x\rightarrow 0
\end{equation}
for all $m_\gamma$ in \eqref{ams}. 

Thus we reach an interesting conclusion. In the limit $x\rightarrow 0$ the only effect on the equations for the eikonal moments compared to the {\bf weak field limit} is a finite renormalization of $j^\mu_{0}$. Equations of motion for $a^\mu_0$ can be of course expanded in Taylor series in $x^+$. This yields {\bf linear} equations of motion for the eikonal moments $a^\mu_{\alpha 0}$, same as in the weak field limit, but with currents that include finite renormalization due to the contribution of high longitudinal momentum modes.

This is good news; it resolves the potential paradox of the instability of the classical equations in the leading eikonal order due to interaction with subeikonal components. One does have to keep in mind though that when one for example prescribes the distribution of color charge density in the McLerran-Venugopalan model, the charge density in question is not that of the "bare" valence charges but rather includes the {\bf classical} effect of high momentum gluon modes.

The procedure described above is reminiscent of the approach one takes in the derivation of JIMWLK equation. There too the high momentum modes contribute to the quantum evolution of the low longitudinal momentum fields via their contribution to the color charge density.
We do not consider quantum evolution in the present paper, rather it is the high longitudinal momentum modes of the classical field that renormalize the color currents. Interestingly we see that already on the classical level this "integration out" is necessary in order to define a sensible eikonal expansion. 

We also note that increasing the energy of the hadron is equivalent to decreasing $\xi$. Therefore a fixed separation scale $\Lambda$ at higher energy corresponds to smaller values of $x$, such that $x\propto 1/E$ similarly to the scaling in the the JIMWLK evolution.

\section{Strong Fields in The Light Cone Gauge}
Many of the standard calculations at high energy are performed in the light cone gauge. We should therefore verify that the conclusions we reached in the previous section hold also in this gauge.

Although we can in principle use the same Lagrangian framework, it turns out that in the light cone gauge it is much less cumbersome to use a Hamiltonian rather than Lagrangian setup. We thus consider the light cone Hamiltonian \cite{BO1} coupled to external sources
\begin{equation}
    H=\int d^3x\left\{\frac{1}{2}\left[\frac{1}{2}F^{ij}_{a}F^{ij}_{a}+F^{+-}_{a}F^{+-}_{a}\right]-A^i_aJ^i_a\right\}
\end{equation}
where in the above we set  $A^{+a}=0$, and the longitudinal component of vector potential is expressed in terms of the transverse components as
\begin{equation}
    A^-_a=\frac{1}{(\partial^+)^2}\left[D^i_{ab}\partial^+A^{ib}-J^{+a}\right]
\end{equation}
As before, our definition of the covariant derivative is
\begin{equation}
    D^i_{ab}\equiv \partial^i\delta_{ab}-gf^{acb}A^i_{c}
\end{equation}
 Although we do not indicate this explicitly, one has to keep in mind that in the Hamiltonian formulation the positive and negative longitudinal momentum modes $A^i_a(p^+,p_\perp)$ are canonically conjugate variables.

As we already know from the consideration of the weak field limit, the eikonal expansion in the light cone gauge is somewhat different than that in the Lorenz gauge. In particular, in the presence of charges the transverse components of the vector potential do not vanish at longitudinal infinity $x^-\rightarrow \infty$. Thus we need to keep the mode proportional to $1/p^+$ in the eikonal expansion of the vector potential in  momentum space. This mode corresponds to the first term in \eqref{lceik}.
 
 Our first step is to split the components  of the vector potential into high and low momentum modes aiming to derive the Hamiltonian for the low momentum fields. This time only the two transverse components of the vector potential are independent and we use the decomposition only for those two. 
 
First, we have to carefully define the inverse of the longitudinal momentum. The definition we use is 
\begin{equation}\label{eps}
    \frac{1}{\partial ^+}f\equiv \int_{-\infty}^{x^-}d y^-f(y^-)
\end{equation}
which with our conventions for Fourier transform, 
\begin{equation}
    f(x)=\int \frac{dp}{2\pi} e^{-ipx}f(p); \ \ \ \ f(p)=\int d xe^{ipx}f(p)
    \end{equation}
translates in momentum space into
\begin{equation}
    \frac{1}{\partial^+}=\frac{i}{p^++i\epsilon}
\end{equation}

Following a procedure analogous to that of the previous section we are tempted to define
the low momentum fields as (we will only explicitly keep $a^i_0$ and $a^i_1$ in this section) 
\begin{equation}\label{lowm}
    A^{ia}_L=\theta(\Lambda-|p^+|)\Big[\frac{i}{p^++i\epsilon} a_0^{ia}+\xi a_1^{ia}+...\Big]
\end{equation}
This however turns out to be problematic because of the nonanalytic behavior of the $\theta$ function in \eqref{lowm}. It is easy to see that when transformed into coordinate space it amounts to regulating the $\delta$-function such that it decays at infinity only very slowly - as $1/x^-$ modulo oscillations. This, combined with non vanishing asymptotic value of vector potentials leads to spurious divergencies in the nonlinear terms in some of the field components constructed from the low momentum fields \eqref{lowm}. 

To avoid this technical issue, it is convenient to somewhat alter the way the field is split into low and high momentum modes. Instead of \eqref{lowm} we will use
\begin{equation}\label{low1}
    A^{ia}_L=\frac{\Lambda^2}{(p^+)^2+\Lambda^2}\frac{1}{2}\cos\frac{p^+}{\Lambda}\Big[\frac{i}{p^++i\epsilon} a_0^{ia}+\xi a_1^{ia}+...\Big]
\end{equation}
This has the advantage
 that
the 
regularized $\delta$-function in coordinate space now behaves as
\begin{equation}
\delta_\Lambda(x^-)\sim \Lambda e^{-\Lambda|x^-|}f(x^-)
\end{equation}
where $f(x)\rightarrow_{|x|\rightarrow \infty} const$. 
This is well behaved at infinity, as it decays faster than any power of $x$ and therefore this definition is free from the problem  mentioned above.

The aim of the introduction of the oscillating factor in \eqref{low1} is to simplify explicit calculations of convolutions of the fields in momentum space by allowing the integration contour to be closed at imaginary infinity. This factor is relevant for calculating convolutions of high enough eikonal components, where the prefactor $\frac{\Lambda^2}{(p^+)^2+\Lambda^2}$ may not be sufficient to ensure the vanishing of the integrand at imaginary infinity.
For the first several eikonal orders however this factor is irrelevant.  Since our explicit calculations below involve only the first two eikonal orders, to make life easier, here we will use the simplified expression
\begin{equation}\label{low}
    A^{ia}_L=\frac{\Lambda^2}{(p^+)^2+\Lambda^2}\Big[\frac{i}{p^++i\epsilon} a_0^{ia}+\xi a_1^{ia}+...\Big]
\end{equation}
and
\begin{equation}
    A_H^{ia}(p^+)=\left[1-\frac{\Lambda^2}{(p^+)^2+\Lambda^2}\right]A^{ia}(p^+)
\end{equation}
With this definition we have
\begin{eqnarray}
    A^{-a}_L&=&\frac{i}{p^++i\epsilon}\Bigg[\frac{\Lambda^2}{(p^+)^2+\Lambda^2}
    \left[\frac{i}{p^++i\epsilon}\partial^ia^{ia}_0+\xi \partial^ia^{ia}_1\right]-[\frac{i}{p^++i\epsilon}j_0^{+a}+\xi j_1^{+a}]\\
    &&\ \ \ \ \ \ \ \ \ \ \ \ \ \ +i\xi \frac{g}{2}\frac{\Lambda^2}{p^++i\Lambda}\frac{i}{p^++i\epsilon}f^{abc}a^{ib}_0a_1^{ic}\Bigg]\nonumber
\end{eqnarray}

We can now write down the low momentum components of the field strength tensor necessary to calculate the Hamiltonian for the low momentum modes. 
These depend on the low momentum components of the vector potential, but also on the high momentum components, which contribute to the nonlinear terms. For $p^+\lesssim \Lambda$ we have

\begin{eqnarray}\label{lclh}
    F^{+-}_L(p^+)&=&\frac{\Lambda^2}{(p^+)^2+\Lambda^2}
    \left[\frac{i}{p^++i\epsilon}\partial^ia^{ia}_0+\xi \partial^ia^{ia}_1\right]-[\frac{i}{p^++i\epsilon}j_0^{+a}+\xi j_1^{+a}]+i\xi\frac{g}{2}\frac{\Lambda^2}{p^++i\Lambda}\frac{i}{p^++i\epsilon}f^{abc}a^{ib}_0a_1^{ic}\nonumber\\ 
    &-&gf^{abc}\frac{i}{p^++i\epsilon}\int_{k^+}A_H^{ib}(p^+-k^+)A_H^{ic}(k^+)\\
     F^{ij}_L(p^+)&=&\frac{\Lambda^2}{(p^+)^2+\Lambda^2}\Big[\frac{i}{p^++i\epsilon} [\partial^ia_0^{ja}-\partial^ja_0^{ia}]+\xi [\partial^ia_1^{ja}-\partial^ja_1^{ia}]\Big]\nonumber\\
    &-&gf^{abc}a_0^{ib}a_0^{jc}\frac{\Lambda^2}{(p^+)^2+\Lambda^2}\left[\frac{i}{p^++i\epsilon}-\frac{3\Lambda}{(p^+)^2+4\Lambda^2}\right]\nonumber\\
    &-&\frac{i\xi g}{2}f^{abc}[a_0^{ib}a_1^{jc}+a_1^{ib}a_0^{jc}]\left[-\frac{\Lambda p^+}{(p^+)^2+4\Lambda^2}+\frac{\Lambda}{p+i\Lambda}\right]-\xi^2 g\frac{\Lambda^2}{(p^+)^2+4\Lambda^2}f^{abc}a^{ib}_1a^{jc}_1\nonumber\\
    &-&gf^{abc}\int_{k^+}A_H^{ib}(p^+-k^+)A_H^{jc}(k^+)\nonumber
    \end{eqnarray}
Using the expressions \eqref{lclh} we can calculate the part of the Hamiltonian that involves the low momentum modes (here we only keep terms up to order $\xi$ in the eikonal expansion in $p^+$):
 \begin{equation}\label{hla}
        \begin{aligned}
            H & =\int_{x_\perp} \Bigg\{\left[\frac{1}{2}\left(\frac{1}{2\epsilon}-\frac{3}{4\Lambda}\right)(\partial^ia^{ai}_0-j^{a+}_{0R})^2
            -\frac{1}{4\Lambda}j^{a+}_{0R}\partial^ia^{ai}_0\right]\\
             & + \left[\frac{1}{4}\left(\frac{1}{2\epsilon}-\frac{3}{4\Lambda}\right)(\partial^ia_0^{aj}-\partial^ja_{0}^{ai}-gf^{abc}a_0^{bi}a_0^{cj})^2+\frac{3}{16\Lambda}gf^{abc}f^{ij a}_0a^{bi}_0a^{cj}_0-\frac{61}{384\Lambda}\left(gf^{abc}a^{bi}_0a^{cj}_0\right)^2\right]\\
            & + \xi\left[\frac{1}{2}(\partial^ia^{ai}_0-j^{a+}_{0R})(\partial^ia^{ai}_1-j^{a+}_{1R}) +\frac{\Lambda}{2}\left(\frac{1}{2\epsilon}-\frac{3}{4\Lambda}\right) gf^{abc}(\partial^ia^{ai}_0-j^{a+}_{0R})a^{bj}_0a^{cj}_1+\frac{1}{8}gf^{abc}j^{a+}_{0R}a^{bj}_0a^{cj}_1\right]\\
            &+\xi\left[\frac{1}{4}(\partial^ia_0^{aj}-\partial^ja_0^{ai})\left[(\partial^ia_1^{aj}-\partial^ja_1^{ai})-\frac{2g}{3}f^{abc}(a^{bi}_0a^{cj}_1+a^{bi}_1a^{cj}_0)\right]-\frac{g}{6}f^{abc}(\partial^ia_1^{aj}-\partial^ja_1^{ai})a^{bi}_0a^{cj}_0\right.\\
            &\left.+\frac{1}{4}g^2f^{abc}f^{ade}a^{bi}_0a^{cj}_0a^{di}_0a^{ej}_1-\frac{1}{2}a^{ai}_0j^{ai}_{0R}\right]\Bigg\}
        \end{aligned}
    \end{equation}
where the "renormalized" color currents are
\begin{eqnarray}\label{j+r}
    j_{0R}^{+a}&=&j_0^{+a}-igf^{abc}\int_{k^+} k^+A_H^{ib}(-k^+)A_H^{ic}(k^+); \\
    \xi j_{1R}^{+a}&=&\xi j_1^{+a}-gf^{abc}\int_{k^+} \left[\frac{\partial}{\partial k^+}\big(-ik^+A_H^{ib}(-k^+)\big)\right]A_H^{ic}(k^+);\nonumber
\end{eqnarray}
and
\begin{eqnarray}\label{jir}
    \xi j^{ai}_{0R}&=&\xi j^{ai}_{0}-gf^{abc}\partial^j\left[\int_{k^+}A_H^{ib}(-k^+)A_H^{jc}(k^+)\right];\\
    \xi j^{ai}_{1R}&=&\xi j^{ai}_{1}+gf^{abc}\partial^j\left[\int_{k^+}\left[\frac{\partial}{\partial k^+}A_H^{ib}(-k^+)\right]A_H^{jc}(k^+)\right]\nonumber
\end{eqnarray}

In these expressions we recognize the features mentioned earlier. For large $\Lambda$ (i.e. of the order of the UV cutoff) for all physical momenta, we have for example
\begin{equation}
F^{+-}_L(p^+)\rightarrow_{\Lambda\rightarrow\infty}
    \left[\frac{i}{p^++i\epsilon}\partial^ia^{ia}_0+\xi \partial^ia^{ia}_1\right]-[\frac{i}{p^++i\epsilon}j_0^{+a}+\xi j_1^{+a}]-\xi\frac{g}{2}\Lambda\frac{i}{p^++i\epsilon}f^{abc}a^{ib}_0a_1^{ic}
    \end{equation}
    The contribution of high momentum modes here vanishes, as all physical momenta are below the cutoff scale $\Lambda$.
This exhibits the mixing of eikonal orders, since the nonlinear term, although formally of order $\xi$, cannot be neglected relative to the leading eikonal term due to the UV divergent factor $\Lambda$. This is also reflected in the Hamiltonian, which contains  terms of order $\xi$ which scale as $\Lambda$ and are therefore UV divergent as $\Lambda\rightarrow\infty$. On the other hand, taking the scaling of $\Lambda$ as in \eqref{scaling} we see that the coefficient of the nonlinear term in $F^{+-}$ for $p^+\lesssim \Lambda$ becomes $x\mu$. For $x\ll 1$, this term can  be treated perturbatively and its presence does not 
invalidate the dominance of the leading eikonal term. The high mode contribution is nonnegligible in this limit, and the role of the high momentum modes now is to "renormalize" the plus component of the color current, as in \eqref{j+r} and also \eqref{jir}.

At small $\Lambda$, at low longitudinal momentum the behavior of the field $A$ should be the same as that of the eikonal solution. Thus its dependence on $k^+$ should factorize from the dependence on transverse coordinates and light cone time. For such a dependence, $A_H^{ib}(-k^+)A_H^{ic}(k^+)=0$. Thus the low longitudinal momentum region should not contribute to the renormalization in  \eqref{j+r} and we expect the renormalization to be finite for $x\rightarrow 0$.

The magnetic field $F^{ij}_L$ is somewhat more regular than the electric field $F^{+-}_L$. It does not exhibit UV divergencies in the infinite $\Lambda$ limit. This is indeed what we expect. Formally the nonlinear $O(\xi)$ contribution is proportional to $\theta(x^-)\delta(x^-)$ and in momentum space is a constant. The value of this constant is ill defined and the purpose of the regulator $\Lambda$ is precisely to define this constant unambiguously in \eqref{lclh}. Our regulator defines this constant at $1/2$, as follows from the value of the coefficient of the $a^i_0a^j_1$ term in \eqref{lclh} at $p^+\rightarrow 0$. At small $\Lambda$ again the contribution of the high momentum modes is important. This contribution directly changes the value of the color magnetic field at low momentum via the last term in \eqref{lclh}, which can be equivalently interpreted as the renormalization of the transverse components of the color current, as in \eqref{hla}. 

One can worry that there might be a potential mismatch in the powers of $\xi$ between the LHS and RHS of \eqref{jir}. However, for small $\Lambda$ the RHS can be written as the low momentum part of $D^iF^{ij}$. Although $F^{ij}$ in the light cone gauge is not a proper tensor, its Lorentz transformation differs from that of a tensor by a phase rotation. Thus we expect that the scaling of its components with respect to a boost is the same as that of a proper vector. In that case the RHS of \eqref{jir} for $x\rightarrow 0$ scales with $\xi$ in the same way as the LHS.

One important point is that the limit $x\rightarrow 0$ has to be taken only after $\epsilon\rightarrow 0$, as $\epsilon$ is an infrared regulator, while the value of $x$ determines a small but finite momentum scale. The leading terms in the $O(1)$ part of the Hamiltonian in \eqref{hla} are therefore of order $1/\epsilon$. Obviously, finiteness of energy requires that the field that multiplies $1/\epsilon$ in the Hamiltoian vanishes on the classical solution. Thus at leading eikonal order we recover the solution used in, for example \cite{mv} 
\begin{eqnarray}
&&    \partial^ia^{ai}_0-j^{a+}_{0R}=0\\
&& \partial^ia_0^{aj}-\partial^ja_{0}^{ai}-gf^{abc}a_0^{bi}a_0^{cj}=0\nonumber
\end{eqnarray}
with the only difference being that the charge density is renormalized by the contribution of high momentum modes.

Thus we conclude that, just like in the Lorenz gauge, the proper way to define the eikonal limit is not in terms of the bare currents, but in terms of renormalized currents that contain the contribution of high momentum modes with the longitudinal momentum $p^+>\Lambda$. For the choice of $\Lambda $ as in \eqref{scaling}
with $x\ll 1$, the current renormalization is finite, and the terms leading to mixing of terms in the eikonal expansion become suppressed by powers of $x$ and can therefore be treated perturbatively.

\section{Conclusions}

In this paper we have presented the formal eikonal expansion in the classical Yang-Mils theory in the context of  looking for solutions to the classical Yang-Mills equations coupled to external currents. The currents  represent the "valence" content of a hadron, e.g. valence quarks that create the classical Yang-Mills fields. We work under assumption that the classical EOM have a well defined solution in the rest frame of a "hadron". By going to a boosted reference frame we have identified the correct expansion in the eikonality parameter $\xi$ with the double expansion of vector potential in powers of the light cone time $x^+$ and in derivatives of the $\delta$-function of the longitudinal coordinate $x^-$ ( i.e. $\delta^{(n)}(x^-)$). This latter expansion is equivalent to Taylor expansion in powers of the longitudinal momentum $p^+$.

This expansion is most straightforward in the covariant gauge, where the components of the vector potential transform under a Lorentz boost as a proper vector. In Light Cone gauge a naive boost transformation has to be accompanied by a gauge transformation in order to preserve the gauge condition in the boosted system. The eikonal  expansion is therefore somewhat different in this gauge. Nevertheless it can be defined, by gauge transforming the problem between the two gauges.

In the weak field limit this expansion is well defined, and the eikonal moments of the color fields are determined directly by the appropriate moments of the currents. In this context we consider the wave function for the fast field modes derived in the Born-Oppenheimer approximation recently in \cite{BO1, BO2}. This wave function is derived in the weak field limit, and
 is a coherent state of the vector potential in light cone gauge. As such, it has a classical interpretation. We show that the expectation value of the vector potential in this state is indeed given by the solution of the classical Yang-Mills EOM. The difference between this solution and the static solution appearing in the JIMWLK approach, is that the former corresponds to a quasi static configuration which does not completely neglect the time dependence of the slow background field modes. Correspondingly, when examined through the prism of the eikonal  expansion, it contains a nonvanishing subeikonal component arising from a slow time dependence of background fields/currents.

 Beyond the weak field limit however, the eikonal expansion taken literally has serious problems. We showed that for strong fields the expansion in eikonality ceases to be well defined. The problem is that nonlinearities in Yang-Mills equations are local in coordinate space. When expanding in derivatives of the $\delta$-function such nonlinear terms are UV divergent and require regularization. The scale of the UV regulator is naturally given by the width of the Lorentz contracted shock wave, which scales as $1/\xi$. Thus the UV divergent terms, which contain higher eikonal moments of the field and are formally higher order in $\xi$, upon regularization mix with lower order eikonal moments in equations of motion. This mixing of orders of the eikonal expansion makes it impossible to include perturbatively higher eikonal moments into a solution of the EOM. Instead one has to solve equations that couple higher moments to lower ones already in leading order, and the concept of expansion disappears.
 
The troubling question then becomes whether this issue threatens  the validity of the leading eikonal approximation which is routinely employed in the high energy limit of hadronic scattering and associated high energy evolution approaches. The way out of the conundrum is to recognize that one indeed cannot employ the eikonal expansion to represent the field at all values of longitudinal momenta. Only at low longitudinal momenta should one be able to approximate the field by a finite number of eikonal terms. Physically therefore the right thing to do is to separate the vector potential into high and low longitudinal momentum modes $A_H$ and $A_L$ respectively. The low momentum field $A_L$ can be approximated by a finite number of terms in the eikonal expansion, while the high momentum modes need to be kept in full. This procedure then leads to equations of motion for $A_L$ in the high momentum field background. The result is that the color charge and current densities which couple to the low eikonal moments that parametrize $A_L$ are "renormalized" as they include contributions due to $A_H$. These equations can indeed be consistently solved using the eikonal expansion for $A_L$. 

The approach introduces an arbitrary momentum scale $\Lambda$, which separates the low and high momentum modes. The situation is remarkably similar to the ubiquitous effective field theory paradigm governing a variety of applications in quantum field theory. Importantly however, here the separation of scales is necessary already on the classical level, before the theory is ever quantized. 

Just like in the effective theory applications, the separation scale cannot be chosen too high. The natural scale of the problem is $\Lambda\sim x \frac{\mu}{\xi}$, where $\mu$ is a typical hadronic scale and $x$ is a number parametrically of order unity. We showed that for large $\Lambda$, i.e. $x\approx 1$ the eikonal order mixing is significant. This is reflected in the coefficient of the effective Lagrangian (or Hamiltonian) for $A_L$. The coefficients of terms that couple higher eikonal moments to lower ones in the Lagrangian become of order unity. Thus one cannot decouple lower eikonal moments from higher ones. On the other hand choosing $\Lambda$ low enough, i.e. $x\ll 1$ makes the couplings suppressed by powers of $x$. One can then consistently forget about higher eikonal moments while solving the equations for the leading eikonal terms, and later include those perturbatively.

The price to pay of course is that the currents that drive the lowest eikonal terms undergo stronger renormalization at lower $x$. Nevertheless this setup allows us to interpret the ambient leading eikonal calculations in a consistent way; one can indeed exclude all eikonal moments except for the leading one from equations of motion if one realizes that the color currents in the equation represent not only the "valence" charges that are simply boosted from the rest frame, but also include contributions of all higher momentum field modes with longitudinal momenta greater than $x\frac{\mu}{\xi}$. 

It is natural to think of the constant $x$ we introduced here as Bjorken $x$. Thus at low $x$ we can indeed determine fields using the leading eikonal approximation, but with fully renormalized currents. One should note that even though we refer to the currents as "renormalized", the renormalization is not at all logarithmic but rather powerlike. This is witnessed by the fact that the coefficients in the effective Lagrangian/Hamiltonian have a power behavior in $x$. Thus this "classical renormalization group" flow is the precursor of the quantum renormalization group approach which leads to quantum JIMWLK evolution equation\cite{jimwlk}. In this sense the logarithmic quantum renormalization of currents of JIMWLK is the quantum correction to the power like renormalization necessary to define a sensible classical eikonal expansion.

So the situation is not all that bad, but there is still a fly in the ointment. Even though the procedure described above allows one to express low eikonal moments of the vector potential in terms of the renormalized currents, this still does not give us the components of field strength at low longitudinal momentum. This is clear from considering for example \eqref{lclh}. Even if we know $A_L$, there is still an additive contribution to $F^{\mu\nu}(p^+)$ at $p^+<\Lambda$ that arises directly from the nonlinear contribution of $A_H$. Such contributions have to be determined independently of $A_L$. Admittedly, these additional terms in the field strength are closely related to the renormalization of the currents in \eqref{j+r} and \eqref{jir}, and it may be possible to determine them if the full information about the currents is available. If this is the case, then specifying the distribution of renormalized currents should be enough to calculate the distribution of the field strength components. This question warrants further investigation.

\section*{Acknowledgement}
This work  is supported by the NSF Nuclear Theory grants 2208387 and 2514546. 
This work is also supported by the U.S. Department of Energy, Office of Science, Office of Nuclear Physics, within the framework of the Saturated Glue (SURGE) Topical Theory Collaboration.
We thank Ming Li and Vladi Skokov for interesting discussions.

\end{document}